\documentclass[aps,PRE,superscriptaddress,showpacs,showkeys,reprint]{revtex4-1}
\usepackage{amsmath}
\usepackage{amsfonts}
\usepackage{amsthm}
\usepackage{graphics}
\usepackage{graphicx}
\usepackage{subfigure}
\usepackage{amssymb}
\usepackage{color}
\usepackage{url}
\usepackage{hyperref}
\usepackage[abs]{overpic}
\usepackage{tikz}
\usepackage{bm}

\begin{document}

\title{Multiscaling in superfluid turbulence: A shell-model study}

\author{Vishwanath Shukla}
\email{research.vishwanath@gmail.com}
\affiliation{Laboratoire de Physique Statistique de l'Ecole Normale 
Sup{\'e}rieure, \\
24 Rue Lhomond, 75231 Paris, France}
\author{Rahul Pandit}
\email{rahul@physics.iisc.ernet.in}
\altaffiliation[\\ also at~]{Jawaharlal Nehru Centre For Advanced
Scientific Research, Jakkur, Bangalore, India.}
\affiliation{Centre for Condensed Matter Theory, Department of Physics, 
Indian Institute of Science, Bangalore 560012, India.} 

\date{\today}
\begin{abstract}
We examine the multiscaling behavior of the normal- and superfluid-velocity
structure functions in three-dimensional superfluid turbulence by using a shell
model for the three-dimensional (3D) Hall-Vinen-Bekharevich-Khalatnikov (HVBK)
equations.  Our 3D-HVBK shell model is based on the Gledzer-Okhitani-Yamada
(GOY) shell model.  We examine the dependence of the multiscaling exponents on
the normal-fluid fraction and the mutual-friction coefficients. Our extensive
study of the 3D-HVBK shell model shows that the multiscaling behavior of the
velocity structure functions in superfluid turbulence is more complicated than
it is in fluid turbulence.
\end{abstract}
\maketitle

\section{Introduction}

The characterization of energy spectra and velocity structure
functions~\cite{frischbook} occupies a central place in the elucidation of the
statistical properties of turbulence, be it in
fluids~\cite{frischbook,meneveau_91,SreenivasanAntonia97,boffetta_08,kanedaannrev,arneodo,donzis_08,pandit_09},
conducting fluids~\cite{biskamp_00,mininni_07,mininni_09,sahoo_11},
or
superfluids~\cite{donnellybook,lathrop2011review,skrbeksreeni2012review,marcpnas14,shukla2014hvbk}.
For example, in fluid turbulence, we often use the longitudinal velocity
$\mathbf{v}$ structure function $S_p(r) \equiv \langle [\delta v(r)]^p
\rangle$, where $\delta v(r) \equiv [\mathbf{v}(\mathbf{x}+\mathbf{r}) -
\mathbf{v}(\mathbf{x})] \cdot [\mathbf{r}/r]$, which scales as $S_p(r) \sim
r^{\zeta_p}$, for $r \equiv \mid \mathbf{r}\mid $ in the inertial range $\eta_d
\ll r \ll L$; viscous dissipation is significant below the dissipation length
scale $\eta_d$; and $L$ is the large length scale at which energy is injected
into the fluid. The exponents $\zeta_p$, which characterize multiscaling, are
nonlinear, monotone increasing functions of $p$~\cite{frischbook}; simple
scaling is obtained if $\zeta_p$ depends linearly on $p$, as in the K41
phenomenological approach of
Kolmogorov~\cite{kolmogorov1941a,kolmogorov1941b,kolmogorov1941c} that yields
$\zeta_p^{K41} = p/3$. 

Direct numerical simulations (DNSs) play an important role in studies of
structure-function multiscaling in fluid
turbulence~\cite{frischbook,kanedaannrev,arneodo}; such DNSs have achieved
impressive spatial resolutions (see, e.g.,
Refs.~\cite{frischbook,kanedaannrev}). By contrast, DNS studies of superfluid
turbulence, whether at the level of the Gross-Pitaevskii (GP) equation
(\cite{vmrnjp13,gpmfric} and references therein) or via the
Hall-Vinen-Bekharevich-Khalatnikov (HVBK) equations
(\cite{shukla2014hvbk,vsthesis} and references therein), have only achieved
modest spatial resolutions. Furthermore, the large number of parameters in these
equations, e.g., the mutual-friction coefficients, the ratio of the
normal-fluid density to the superfluid density, and the Reynolds number, pose a
significant challenge for systematic studies of the multiscaling of
normal-fluid- and superfluid-velocity structure functions. It has been
suggested, therefore, that shell models for the three-dimensional (3D) HVBK
equations~\cite{wacks2011shellmodel,procacciashell3He,
procacciaintermittencyshellmodel} be used first to study such multiscaling in
detail. 

Ever since their introduction in the early work of Obukhov~\cite{obukhov},
Desnyansky and Novikov~\cite{desnovikov}, and Gledzer, and Ohkitani and
Yamada~\cite{goy1,goy2} (henceforth GOY), shell models have played valuable
roles in elucidating the multiscaling properties of structure functions of
fluid turbulence~\cite{frischbook,jensen91,pisarenko93,DharPraman97,DharPRL97,
biferaleshell03,bohrbook,ditlevsenbook,sabrashell}.  Over the years, such shell
models have been used to study magnetohydrodynamic (MHD)
turbulence~\cite{mhdshell1,mhdshell2,mhdshell3,moremhdshell1,moremhdshell2,
moremhdshell3}, Hall-MHD
turbulence~\cite{hallmhdshell1,hallmhdshell2,hallmhdshell3, hallmhdshell4},
fluid turbulence with polymer additives~\cite{polymershell}, fluid turbulence
in two dimensions~\cite{2Dshellmodel}, fluid turbulence in dimensions in
between two and three~\cite{2D3Dshellmodel}, turbulence in binary-fluid
mixtures~\cite{symmbinfluid} and in rotating systems~\cite{rotatingturbulence},
and, as we have mentioned above, turbulence in
superfluids~\cite{wacks2011shellmodel,procacciashell3He,procacciaintermittencyshellmodel}.
Shell models have also been used to initiate studies of the dynamic
multiscaling of time-dependent structure
functions~\cite{dynamicshell1,dynamicshell2,dynamicshell3}. 

We build on the shell-model studies of
Refs.~\cite{wacks2011shellmodel,procacciashell3He,procacciaintermittencyshellmodel}
to explore the dependence of the multiscaling exponents here on the parameters
of the 3D HVBK model. It has been noted in
Ref.~\cite{procacciaintermittencyshellmodel} that, given current computational
resources, a systematic study of this parameter dependence lies beyond the
scope of a well-resolved DNS of the 3D HVBK equations; however, such a study is
possible if we use shell models for these equations.  Our study extends the
work of Refs.~\cite{wacks2011shellmodel,procacciashell3He,
procacciaintermittencyshellmodel} by obtaining a variety of results, which we
summarize, before we present the details of our study.  

Our study of the 3D-HVBK shell model shows that the multiscaling behavior of
the shell-model counterparts of velocity structure functions in superfluid
turbulence is more rich than that reported in
Ref.~\cite{procacciaintermittencyshellmodel}. Our results agree with those of
Ref.~\cite{procacciaintermittencyshellmodel} qualitatively insofar as we find
that, in the limits when the normal-fluid fraction is either small (pure
superfluid) or large, the equal-time multiscaling exponents are close to their
classical-fluid-turbulence values.  In addition, we find that there are two regions, with
intermediate values of the normal-fluid fraction, in which the multiscaling
exponents are larger than those observed for the classical-fluid-turbulence or even
Kolmogorov's 1941 (K41) predictions~\cite{frischbook}; between these two
regions there is a region in which the multiscaling exponents are close to
their K41 values. We have also investigated the dependence of the multiscaling
exponents on the mutual-friction coefficient, with equal proportions of
superfluid and normal-fluid components; here, our results show that, for small
(weak-coupling limit) and large (strong-coupling limit) values of the
mutual-friction coefficient, the multiscaling exponents tend to their
classical-fluid-turbulence values, whereas, in an intermediate range, there are deviations from
the classical-fluid-turbulence behavior; in particular, the multiscaling exponents are larger
than their classical-fluid-turbulence counterparts for high-order structure functions (order
$p\geq3$).

The remainder of this paper is organized as follows. In
Sec.~\ref{sec:model:ch6} we describe the shell model and the numerical methods
we use. Section~\ref{sec:results:ch6} is devoted to our results. We end with
conclusions in Sec.~\ref{sec:conclusions:ch6}.

\section{Models and Numerical Simulations}
\label{sec:model:ch6}

The GOY shell model, for the 3D Navier-Stokes equation~\cite{goy1,goy2} 
for a fluid, comprises the following ordinary differential
equations (ODEs):
\begin{equation}
\begin{split}
\left[\frac{d}{dt} + \nu k_m^2\right]u_m &= \imath
[ak_mu_{m+1}u_{m+2} + bk_{m-1}u_{m-1}u_{m+1} \\
&\quad + ck_{m-2}u_{m-1}u_{m-2}]^*
+ f_m;
\end{split}
\end{equation}

here we label shells by the positive integers $m$, in a logarithmically
discretized Fourier space, with scalar wave numbers $k_m=k_0\lambda^m$, where
$k_0=2^{-4}$ and $\lambda = 2$. The $*$ denotes complex conjugation, $\nu$ is
the kinematic viscosity, and $u_m(k_m)$ are the complex, scalar, shell
velocities. The coefficients $a=1$, $b=-\delta$, $c=-(1-\delta)$ are chosen to
conserve the shell-model analogs of energy and helicity in the limit of
vanishing viscosity and the absence of external forcing; the standard value of
$\delta$ is $1/2$; $N$ is the total number of shells; and $f_m$ is the external
forcing, which is used to drive the system into a turbulent state that is
statistically steady.  The logarithmic discretization of Fourier space allows
us to achieve very high Reynolds numbers, even with a moderate number of
shells. In the GOY-shell-model equations, direct interactions are limited to
the nearest- and next-nearest-neighbor shells; in contrast, if we write the
Navier-Stokes equation in Fourier space, every Fourier mode of the velocity is
directly coupled to every other Fourier mode.

The simplest form of the incompressible, 3D HVBK 
equations~\cite{donnellybook,roche2009HVBK3d} is
\begin{subequations}
\begin{align}
\rho_s\frac{D \mathbf{u}^s}{D t}
&=-\frac{\rho_s}{\rho}\nabla p + \rho_s\sigma\nabla T+\mathbf{F}^s_{mf}, 
\label{eq:HVBK3dus} \\
\rho_n\frac{D \mathbf{u}^n}{D t} 
&=-\frac{\rho_n}{\rho}\nabla p - \rho_n\sigma\nabla T+\mathbf{F}^n_{mf}
+ \nu_n\nabla^2\mathbf{u}^n, \label{eq:HVBK3dun}
\end{align}
\end{subequations}
with $D\mathbf{u}^i/Dt=\partial \mathbf{u}^i/\partial
t+\mathbf{u}^i\cdot\nabla\mathbf{u}^i$, the incompressibility condition
$\nabla\cdot\mathbf{u}^i=0$, and the superscript $i\in(n,s)$ denotes the normal
fluid ($n$) or the superfluid ($s$); $p$, $\sigma$, and $T$ are the pressure,
specific entropy, and temperature, respectively;  $\rho_n$ ($\rho_s$) is the
normal-fluid (superfluid) density; $\nu_n$ is the kinematic viscosity of the
normal fluid.  The mutual-friction terms, which model the interaction between
the normal and superfluid components, can be written as
$\mathbf{F}^s_{mf}=-(\rho_n/\rho)\mathbf{f}_{mf}$ and
$\mathbf{F}^n_{mf}=(\rho_s/\rho)\mathbf{f}_{mf}$ in Eqs.~(\ref{eq:HVBK3dus})
and (\ref{eq:HVBK3dus}), respectively, where
\begin{equation}\label{eq:mf:ch6}
\mathbf{f}_{\rm mf} = \frac{B}{2}\frac{\mathbf{\omega}_{\rm s}}{|\mathbf{\omega}_{\rm s}|}\times
(\mathbf{\omega}_{\rm s}\times\mathbf{u}_{\rm ns})
+ \frac{B'}{2}\mathbf{\omega}_{\rm s}\times\mathbf{u}_{\rm ns},
\end{equation} 
with $\mathbf{u}_{\rm ns} = (\mathbf{u}_{\rm n} - \mathbf{u}_{\rm s})$ the slip
velocity, and $B$ and $B'$ the coefficients of mutual friction.  In most of our
studies we set $B'=0$, so $\mathbf{f}_{\rm mf} = -\frac{B}{2}|\omega_{\rm
s}|\mathbf{u}_{\rm ns}$, which is the Gorter-Mellink
form~\cite{gortermellink1949}.

We use the following shell model for the 3D HVBK equations; it is
based on the GOY shell model for a fluid~\cite{wacks2011shellmodel}.
\begin{equation}\label{eq:goyn}
\left[\frac{d}{dt} + \nu_n k_m^2\right]u^n_m = {\rm NL}[u^n_m] + F^n_m + f^n_m,
\end{equation}
\begin{equation}\label{eq:goys}
\left[\frac{d}{dt} + \nu_s k_m^2\right]u^s_m = {\rm NL}[u^s_m] + F^s_m + f^s_m,
\end{equation}
where
\begin{equation}
\begin{split}
{\rm NL}[u_m] &= \imath [ak_mu_{m+1}u_{m+2} + bk_{m-1}u_{m-1}u_{m+1} \\
&\quad + ck_{m-2}u_{m-1}u_{m-2}]^*.
\end{split}
\end{equation}
Here, as in the GOY model, we have a logarithmically discretized Fourier space
with shell-$m$ wave numbers $k=k_0\lambda^m$, where $k_0=2^{-4}$ and
$\lambda=2$, and kinematic viscosities $\nu_n$ and $\nu_s$ for the normal fluid
and the superfluid, respectively; of course, $\nu_s$ must vanish in a
superfluid but, in practical numerical simulations, $\nu_n \gg \nu_s > 0$ for
numerical stability. The normal and superfluid dynamical variables are,
respectively, the complex, scalar, shell velocities $u^n_m(k_m)$ and
$u^s_m(k_m)$; and $f^n_m$ and $f^s_m$ are the external forcing terms. The
coefficients  $a=1$, $b=-1/2$, $c=-1/2$ are chosen to conserve the shell-model
analogs of energy and helicity in the limit of vanishing viscosity and the
absence of external forcing.  The shell-model analogs of the mutual-friction
terms, which models the interaction between the normal and the superfluid
components, are
\begin{equation}
F^s_m = \frac{\rho_nB\Omega^{1/2}_s}{2\rho}(u^n_m - u^s_m)
\end{equation}
and
\begin{equation}
F^n_m = -\frac{\rho_sB\Omega^{1/2}_s}{2\rho}(u^n_m - u^s_m).
\end{equation}
The shell-model superfluid and normal-fluid enstrophies are, respectively, 
\begin{equation}
\Omega_s = \sum^N_{m=1} \frac{1}{2}k_m^{2}|u^s_m|^2
\end{equation}
and
\begin{equation}
\Omega_n = \sum^N_{m=1} \frac{1}{2}k_m^{2}|u^n_m|^2.
\end{equation}
The total energy is
\begin{equation}
E_T = E_n + E_s\equiv\frac{1}{2}\sum^N_{m=1}\bigl(|u^n_m|^2+|u^s_m|^2\bigr),
\end{equation}
where $E_n$ and $E_s$ are the normal-fluid and superfluid energies,
respectively. Other statistical quantities that we use in our study are as
follows: The helicity is
\begin{equation}
H_i = \sum^N_{m=1} \frac{1}{2}\Bigl(\frac{a}{c}\Bigr)^m\frac{|u^i_m|^2}{k_m};
\end{equation}
the energy spectra are
\begin{equation}
E_{i}(k_m)=\frac{1}{2}\frac{|u^i_m|^2}{k_m};
\end{equation}
the root-mean-square velocities are
\begin{equation}
u^{i}_{\rm rms} = \Bigl(\sum_m|u^i_m|^2\Bigr)^{1/2};
\end{equation}
the Taylor microscale is
\begin{equation}
\lambda_i=\Biggl[\frac{\sum_m E^i(k_m)}{\sum_m k^2_m E^i(k_m)}\Biggr]^{1/2};
\end{equation}
the Taylor-microscale Reynolds number is
\begin{equation}
Re^i_{\lambda} = u_{\rm rms}\lambda_i/\nu_i;
\end{equation}
the integral length scale is
\begin{equation}
\ell_{I} = \frac{\sum_m E^i(k_m)/k_m}{\sum_m E^i(k_m)};
\end{equation}
and the large-eddy-turnover time is
\begin{equation}
T^i_{eddy} = \frac{1}{k_1u^i_1};
\end{equation}
here and henceforth $i\in (n,s)$.

The equal-time, order-$p$ structure functions for the shell model are
\begin{equation}\label{eq:goysf}
S^i_p(k_m) \equiv \Bigl<\bigl[u^i_m(t)u^{i*}_m(t)\bigr]^{p/2}\Bigr> \sim k_m^{-\zeta^i_p},
\end{equation}
where the power-law dependence is obtained only if $k^{-1}_m$ lies in the
inertial range. The structure functions defined above show period-three
oscillations because of three cycles in the static solutions of the GOY model
for the Navier-Stokes equation~\cite{DharPraman97}.  Therefore, we use the
modified structure functions~\cite{pisarenko93,DharPraman97}
\begin{equation}\label{eq:strsigma}
\Sigma^i_p \equiv \Bigl< \left\vert\Im\bigl[u^i_{m+2}u^i_{m+1}u^i_{m}
-\frac{1}{4}u^i_{m-1}u^i_{m}u^i_{m+1}\bigr]\right\vert^{p/3}\Bigr>
\sim k_m^{-\zeta^i_p},
\end{equation}
which filter out these oscillations effectively.  The Sabra-model
variant~\cite{procacciashell3He,procacciaintermittencyshellmodel} of the 3D
HVBK equations does not show such oscillations.  
We expect that the multiscaling exponents
$\zeta^i_p$, $i\in (n,s)$, satisfy the following convexity inequality for
any three positive integers $p_1\leq p_2\leq p_3$~\cite{frischbook}:
\begin{equation}\label{eq:convexchap6}
(p_3-p_1)\zeta^i_{2p_2}\geq (p_3-p_2)\zeta^i_{2p_1} + (p_2-p_1)\zeta^i_{2p_3}.
\end{equation}
We obtain smooth energy spectra, without period-$3$ oscillations, by using
$E_i(k_m)=\Sigma^i_2(k_m)/k_m$, $i\in (n,s)$.

To obtain a turbulent, but statistically steady, state, we force both the
superfluid and the normal-fluid components with the forces 
\begin{equation}
f^{n,s}_m = (1+\imath)\times5\times10^{-3}\delta_{1,m}, 
\end{equation} 
where $\delta_{1,m}$ is the Kronecker delta.  We use the second-order, slaved
Adams-Bashforth scheme to integrate the 3D-HVBK-shell-model
Eqs.~(\ref{eq:goyn}) and (\ref{eq:goys})~\cite{pisarenko93,cox}.  To study the
multiscaling behaviors of structure functions here, we design the following
three sets of runs: 
\begin{enumerate}
\item $\tt G1a$-$\tt G9$: In these runs, we use the values of $\rho_n/\rho$ and
$B$, which have been measured at different temperatures in experiments on
helium II~\cite{donnelly1998omfdata}. We use suitable values of $\nu_{n}$ and
$\nu_s$, which we list, along with other parameters, in
Table~\ref{table:paramexpt}.  
\item $\tt B1$-$\tt B19$: We vary $\rho_n/\rho$ between $0.05-0.95$ and keep
$B=1.5$ fixed.
\item $\tt R1$-$\tt R12$: We vary $B$ between $0.1-10$ and keep
$\rho_n/\rho=0.5$ fixed.
\end{enumerate}

In the runs $\tt B1$-$\tt B19$ and $\tt R1$-$\tt R12$, we use $\nu_n=10^{-7}$,
$\nu_s=10^{-9}$, and the time step $\Delta t=10^{-5}$.

We use the initial condition $u^{n,s}_m = (1+\imath)k_me^{-k_m^2}$, for $1 \leq m
\leq N$, in the runs $\tt G1a$-$\tt G9$, $\tt PG1$, and $\tt PG2$; the
GOY-shell-model runs $\tt PG1$ ($\nu_n=10^{-7}$) and $\tt PG2$
($\nu_n=10^{-9}$) are included for the purpose of comparison with the runs $\tt
G1a$-$\tt G9$.  In the runs $\tt B1$-$\tt B19$ and $\tt R1$-$\tt R12$, we use
the initial values $u^{n,s}_m = u^{n,s}_0k_m^{1/2}e^{-k_m^2}e^{i\vartheta_m}$,
for $1\leq m \leq N$, where $\vartheta_m$ is a random phase distributed
uniformly on $[0,2\pi)$. We use the boundary conditions
$u^i_{-2}=u^i_{-1}=u^i_0=0$ and $u^i_{N+1}=u^i_{N+2}=0$, $i\in (n,s)$.  We
report results for $N=36$ shells; Ref.~\cite{wacks2011shellmodel} uses $N=18$
and Ref.~\cite{procacciaintermittencyshellmodel} presents data with $N=36$.

\begin{table}
\begin{center}
\small
   \begin{tabular}{@{\extracolsep{\fill}} c c c c c c c c }
    \hline
    $ $ & $\rho_n/\rho$ & $B$ & $\nu_n$ & $\nu_s$  & $\Delta t$\\ 
   \hline \hline
    {\tt PG1}  &  $-$ &  $-$	& $10^{-7}$ &  $-$ & $10^{-5}$ \\
    {\tt PG2}  &  $-$ &  $-$	& $10^{-9}$ &  $-$ & $10^{-5}$\\
    {\tt G1a}  &  $0.0450$ &  $1.5260$	& $10^{-7}$ &  $10^{-10}$ & $5.0\times10^{-6}$\\
    {\tt G1}  &  $0.0450$ &  $1.5260$	& $10^{-7}$ &  $10^{-9}$ & $10^{-5}$\\
    {\tt G2}  &  $0.0998$ &  $1.3255$	& $10^{-7}$ &  $10^{-9}$ & $10^{-5}$\\
    {\tt G3}  &  $0.2503$ &  $1.0765$	& $10^{-7}$ &  $10^{-9}$ & $10^{-5}$\\
    {\tt G4}  &  $0.4004$ &  $0.9838$	& $10^{-7}$ &  $10^{-9}$ & $10^{-5}$\\
    {\tt G5}  &  $0.4994$ &  $0.9848$	& $10^{-7}$ &  $10^{-9}$ & $10^{-5}$\\
    {\tt G6}  &  $0.6003$ &  $1.0447$	& $10^{-7}$ &  $10^{-9}$ & $10^{-5}$\\
    {\tt G7}  &  $0.6493$ &  $1.1034$	& $10^{-7}$ &  $10^{-9}$ & $10^{-5}$\\
    {\tt G8}  &  $0.6995$ &  $1.1924$	& $10^{-7}$ &  $10^{-9}$ & $10^{-5}$\\
    {\tt G9}  &  $0.7501$ &  $1.3267$	& $10^{-7}$ &  $10^{-9}$ & $10^{-5}$\\

\hline
\end{tabular}
\end{center}
\caption{\small  Parameters for our 3D-shell-model runs (classical-fluid-turbulence) $\tt PG1$,
$\tt PG2$ and 3D-HVBK-shell-model runs $\tt G1a$-$\tt G9$: $\rho_n/\rho$ is the 
normal-fluid density fraction; $B$ is the mutual-friction coefficient; 
$\nu_n$ ($\nu_s$) is the normal-fluid (superfluid) viscosity; $\Delta t$ is 
the time step; we use $N=36$ shells in our simulations.}
\label{table:paramexpt}
\end{table} 

\section{Results}
\label{sec:results:ch6}

We now present the results of our study of superfluid and normal-fluid
turbulence in the 3D-HVBK shell-model. We begin with energy spectra and then
examine the parameter dependence of the exponents that characterize the
multiscaling of structure functions.

\begin{figure}
\resizebox{\linewidth}{!}{
\includegraphics[height=12.cm]{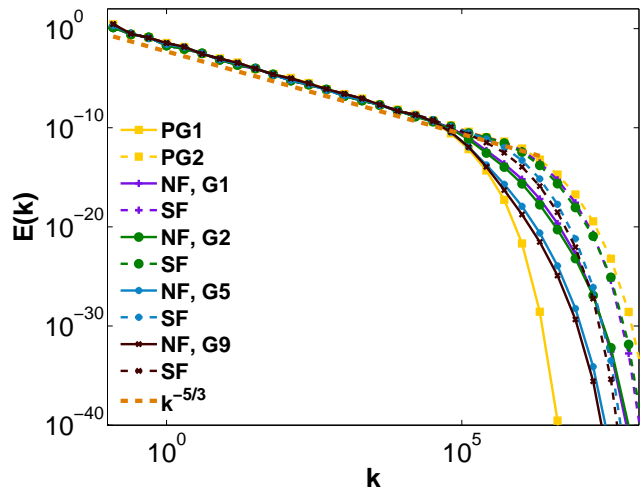}
}
\caption{\small Log-log (base 10) plots of the spectra $E_n(k_m)$ (full curves)
and $E_s(k_m)$ (dashed curves) from our shell-model runs: $\tt PG1$ and $\tt
PG2$ (yellow curves); $\tt G1$ (purple curves); $\tt G2$ (green curves); $\tt
G5$ (sky-blue curves); $\tt G9$ (brown curves); a $k^{-5/3}$ power law is shown
by the orange-dashed line; NF (SF) stands for normal-fluid (superfluid).  }
\label{fig:spectraGruns} 
\end{figure}

In Table~\ref{table:Grunsresults} we list the values of $\lambda_{i}$,
$Re^{i}_{\lambda}$, and $T^{i}_{eddy}$ that we obtain from our
3D-HVBK-shell-model simulations $\tt PG1$, $\tt PG2$, and $\tt G1a$-$\tt G9$.
Figure~\ref{fig:spectraGruns} compares $E_n(k_m)$ (full curves) and $E_s(k_m)$
(dashed curves) for four representative values of $\rho_n/\rho$ (runs $\tt G1$
(purple curves), $\tt G2$ (green curves), $\tt G5$ (sky-blue curves), and $\tt
G9$ (brown curves)). The inertial ranges of $E_n(k_m)$ and $E_s(k_m)$ exhibit
scaling that is consistent with a $k^{-5/3}$ power-law form (orange, dashed
line); of course, this exponent is not exactly $-5/3$ if the structure
functions display multiscaling.  The runs $\tt PG1$ and $\tt PG2$ can be
regarded as uncoupled ($B=0$) normal fluid and superfluid, respectively; we use
them for the sake of comparison with other runs to show how the mutual friction
modifies the energy spectra.  When we couple the normal and superfluid
components, as in the run $\tt G1$, $E_n(k_m)$ is pulled up towards $E_s(k_m)$,
by virtue of the mutual-friction-induced tendency of locking between $u_n$ and
$u_s$ (see Ref.~\cite{shukla2014hvbk}); in contrast, in the absence of
coupling, the spectra $E(k_m)$ for the runs $\tt PG1$ (yellow, full curves) and
$\tt PG2$ (yellow, dashed curves) lie far apart, especially in the dissipation
range.

\begin{table}
\begin{center}
\small
   \begin{tabular}{@{\extracolsep{\fill}} c c c c c c c c }
    \hline
    $ $ & $\lambda_n$ & $\lambda_s$ & $Re^n_{\lambda}(\times10^6)$ &
    $Re^s_{\lambda}(\times10^8)$  & $T^n_{eddy}$ & $T^s_{eddy}$\\ 
   \hline \hline
    {\tt PG1}  &  $0.95$ &  $-$	& $7.2$ &      $-$ & $14.50$ & $-$\\
    {\tt PG2}  &  $0.50$ &  $-$	& $310$  & $-$ & $20.33$ & $-$\\
    {\tt G1a}  &  $0.42$ & $0.28$	& $2.3$ &  $15$ & $45.61$ & $45.61$\\
    {\tt G1}  &  $0.70$ &  $0.51$	& $4.2$ &  $3.1$ & $21.80$ & $21.80$\\
    {\tt G2}  &  $0.73$ &  $0.54$	& $4.4$ &  $3.2$ & $22.03$ & $22.03$\\
    {\tt G3}  &  $0.82$ &  $0.61$	& $5.2$ &  $3.9$ & $19.83$ & $19.83$\\
    {\tt G4}  &  $0.71$ &  $0.54$	& $4.4$ &  $3.4$ & $18.41$ & $18.41$\\
    {\tt G5}  &  $0.89$ &  $0.70$	& $6.2$ &  $4.9$ & $17.15$ & $17.15$\\
    {\tt G6}  &  $0.94$ &  $0.77$	& $6.9$ &  $5.6$ & $15.45$ & $15.45$\\
    {\tt G7}  &  $0.94$ &  $0.78$	& $7.0$ &  $5.9$ & $14.94$ & $14.94$\\
    {\tt G8}  &  $0.95$ &  $0.80$	& $7.2$ &  $6.1$ & $14.51$ & $14.51$\\
    {\tt G9}  &  $0.95$ &  $0.82$	& $7.3$ &  $6.3$ & $14.42$ & $14.42$\\

    {\tt }  &  $$ &  $$	& $$ &  $$ & $$ & $$\\
\hline
\end{tabular}
\end{center}
\caption{\small Parameters from our shell-model runs $\tt PG1$, $\tt
PG2$, and $\tt G1a$-$\tt G9$: $\lambda_n$ ($\lambda_s$) is the Taylor microscale
for the normal-fluid (superfluid); $Re^n_{\lambda}$ ($Re^s_{\lambda}$) is the
Taylor-microscale Reynolds number for the normal-fluid (superfluid);
$T^n_{eddy}$ ($T^s_{eddy}$) is the large-eddy-turnover time for the
normal-fluid (superfluid).}
\label{table:Grunsresults}
\end{table}

We study the multiscaling behaviors of the velocity structure functions for the
3D-HVBK shell-model by calculating the multiscaling exponents $\zeta^n_p$ and
$\zeta^s_p$, for the normal fluid and superfluid components, respectively, by
using the Eqs.~(\ref{eq:strsigma}) for $\Sigma^i_p$. In
Table~\ref{table:zetapGruns} in the Supplemental Material, we list the values
of these exponents, which we have obtained from $\Sigma^i_p$, for $p=1$ to $6$,
$i\in (n,s)$; each row of this Table has two lines; the first and second lines
contain, respectively, the values of $\zeta^n_p$ and $\zeta^s_p$.
Table~\ref{table:zetapGruns} (Supplemental Material) shows that
$\zeta^n_p=\zeta^s_p$, for $p=1$ to $6$, for the runs $\tt G1$-$\tt G9$,
because of the mutual-friction-induced locking of the normal fluid and
superfluid velocities in the inertial range.

\begin{figure}
\resizebox{\linewidth}{!}{
\includegraphics[height=6.cm]{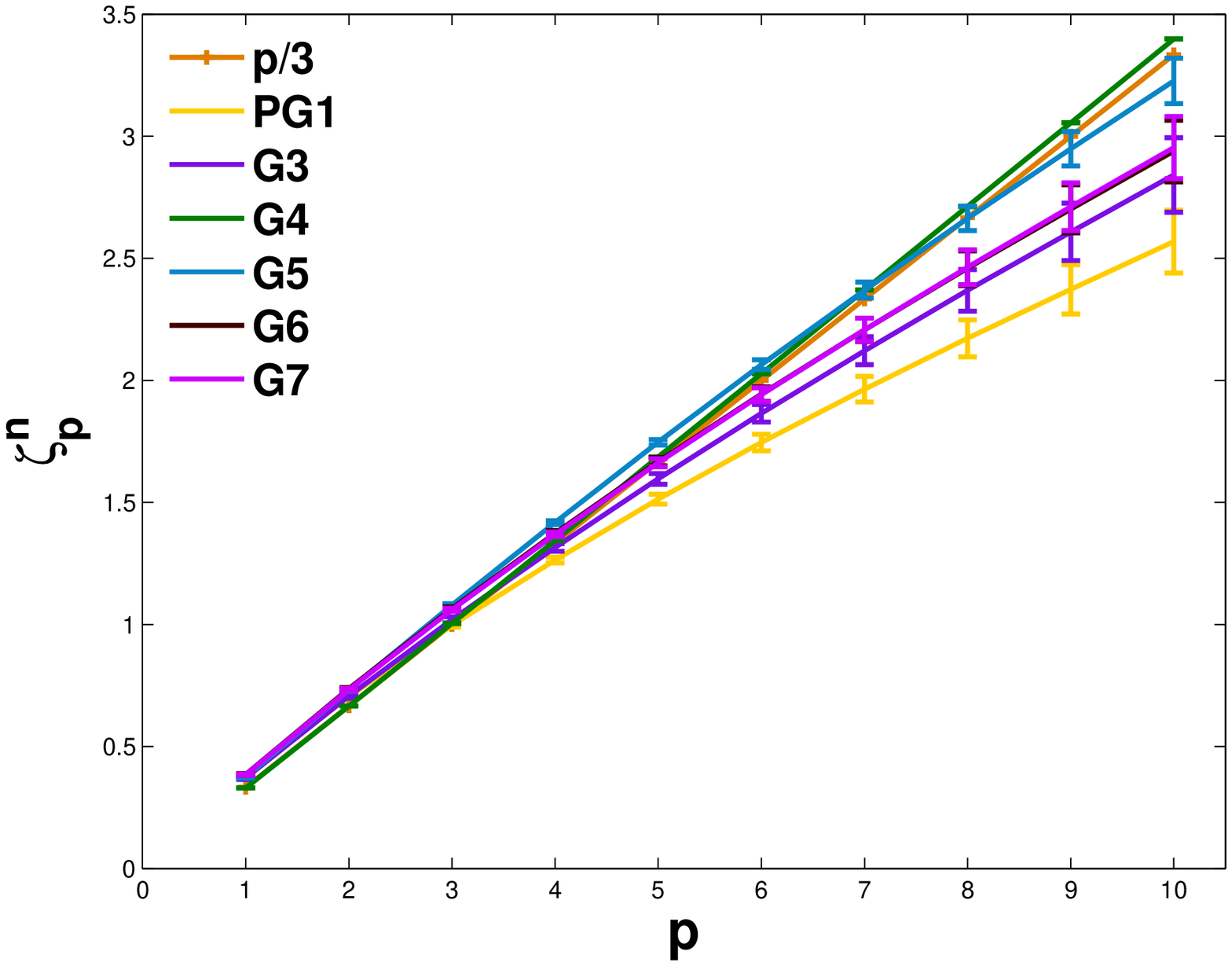}}
\put(-100,30){\bf (a)}
\\
\resizebox{\linewidth}{!}{
\includegraphics[height=6.cm]{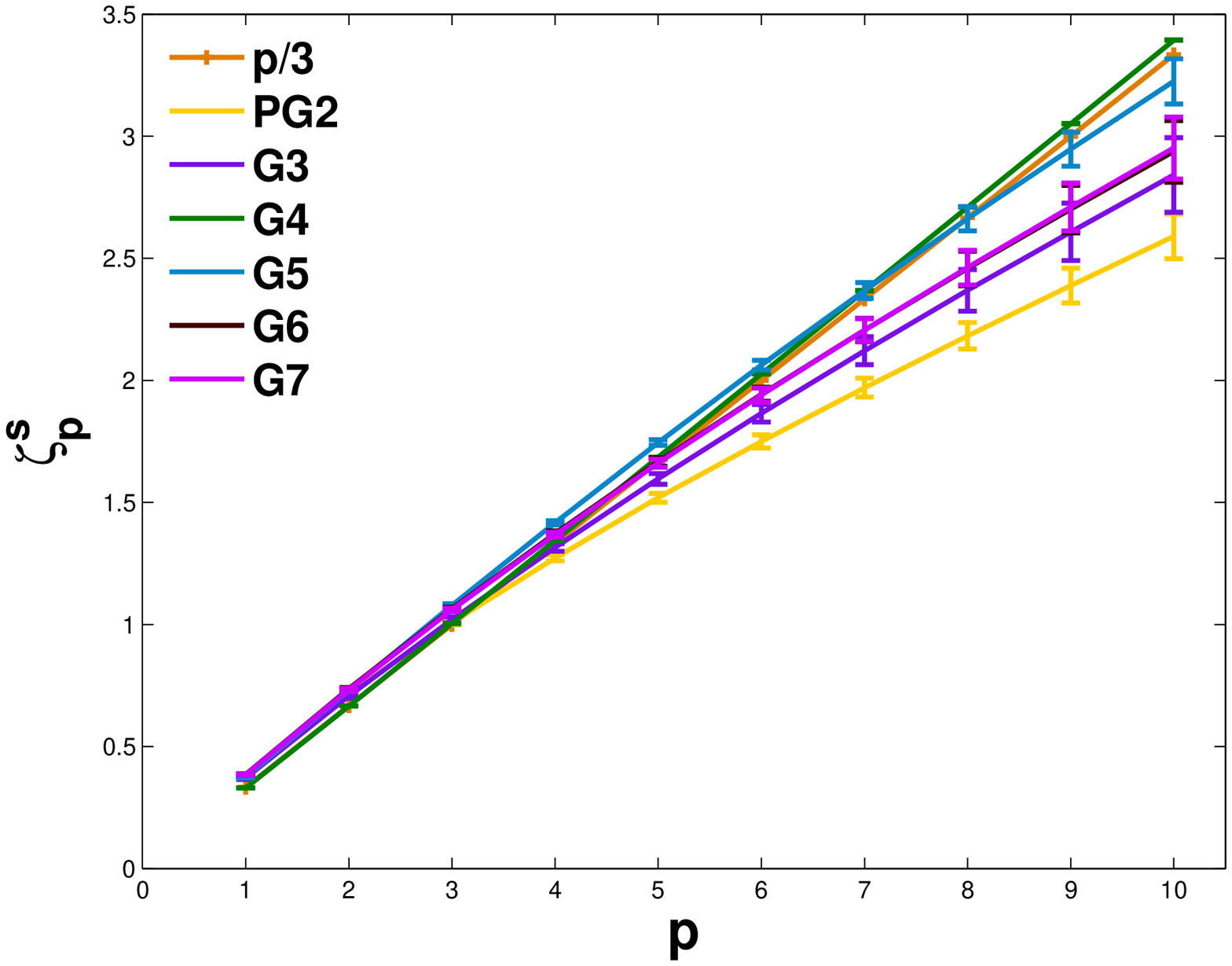}
\put(-100,30){\bf (b)}
}
\caption{\small Plots versus order $p$ of the multiscaling exponents: (a)
$\zeta^n_p$ and (b) $\zeta^s_p$, for the shell-model runs $\tt G3$ (purple
curve), $\tt G4$ (green curve), $\tt G5$ (sky-blue curve), $\tt G6$ (brown
curve), and $\tt G7$ (magenta curve). $\tt PG1$, $\tt PG2$ are the classical-fluid-turbulence
runs (yellow curve) and $\zeta^{K41}=p/3$ is denoted by the orange line.  }
\label{fig:zetapGruns} 
\end{figure}

Figures~\ref{fig:zetapGruns}~(a) and (b) show, respectively, plots of
$\zeta^n_p$ and $\zeta^s_p$ versus the order $p$; in these plots the orange
line is the K41 prediction $\zeta^{K41}_p=p/3$ and the yellow line shows the
multiscaling exponents $\zeta^{c}_p$ of classical (superscript $c$), 3D-fluid
turbulence.  The multiscaling exponents $\zeta^i_p$, $i\in (n,s)$, which we
determine from the 3D-HVBK shell-model, show deviations from $\zeta^{c}_p$;
these deviations depend on the values of $\rho_n/\rho$ and $B$.  Moreover, for
the run $\tt G4$ ($\rho_n/\rho=0.4$, $B=0.9838$), the $\zeta^i_p$'s (green
lines in Figs.~\ref{fig:zetapGruns}~(a) and (b)) are close to
$\zeta^{K41}_p=p/3$.  For the run $\tt G3$ ($\rho_n/\rho=0.25$, $B=1.08$), the
$\zeta^i_p$'s (purple lines in Figs.~\ref{fig:zetapGruns}~(a) and (b)) lie
roughly between $\zeta^{K41}_p$ and $\zeta^c_p$; for the runs $\tt G5$-$\tt
G7$, the differences between $\zeta^i_p$ and $\zeta^{K41}_p$ and $\zeta^c_p$
depend on $p$.

\begin{figure}
\resizebox{\linewidth}{!}{
\includegraphics[height=12.cm]{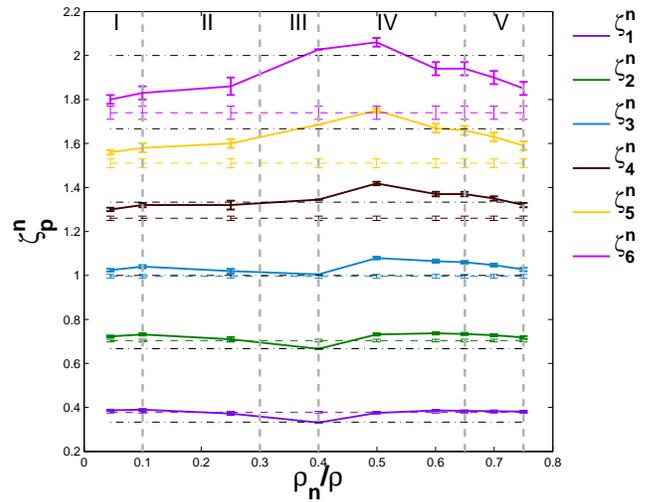}
}
\caption{\small Plots of $\zeta_p^n$, for $p=1$ to $6$, versus $\rho_n/\rho$,
from our shell-model runs $\tt G1$-$\tt G9$.  For the purpose of reference, we
show the value of a classical-fluid-turbulence exponent $\zeta^c_p$, for order $p$, by a
horizontal, dashed line; different colors indicate different values of the
order $p$.  The black, dot-dashed lines indicates $\zeta^{K41}_p=p/3$.} 
\label{fig:zetadndGruns} 
\end{figure}

To understand the dependence of the multiscaling exponents $\zeta^i_p$, $i\in
(n,s)$, on $\rho_n/\rho$ (which includes the variation of $B$ with
temperature), we plot, in Fig.~\ref{fig:zetadndGruns}, $\zeta^n_p$, for $p=1$
to $6$,  versus $\rho_n/\rho$ from our runs $\tt G1$-$\tt G9$.
Figure~\ref{fig:zetadndGruns} shows that, depending on the values of
$\rho_n/\rho$, the behavior of the exponents $\zeta^n_p$ can be classified
{\textit{roughly}} into five regions ${\rm I}$-${\rm V}$ (demarcated by
grey, dashed, vertical lines on the plot). Region ${\rm I}$
($\rho_n/\rho\lesssim 0.1$): The values of $\zeta^n_p$ are close to the
classical-fluid-turbulence exponents $\zeta^c_p$.  Region ${\rm II}$
($0.1<\rho_n/\rho< 0.3$): $\zeta^n_p>\zeta^c_p$, for $p\geq3$ and, for $p=1,2$,
$\zeta^n_p\simeq\zeta^c_p$.  
Region ${\rm III}$ ($0.3\lesssim\rho_n/\rho\lesssim 0.4$): 
$\zeta^n_p\simeq\zeta^{K41}_p$.
Region ${\rm IV}$ ($0.4<\rho_n/\rho\lesssim
0.65$): $\zeta^n_p$ show significant deviations from both $\zeta^c_p$ and
$\zeta^{K41}_p$.  Region ${\rm V}$ ($\rho_n/\rho>0.65$): $\zeta^n_p$ show a
tendency to move towards $\zeta^c_n$.

\begin{figure}
\resizebox{\linewidth}{!}{
\includegraphics[height=12.cm]{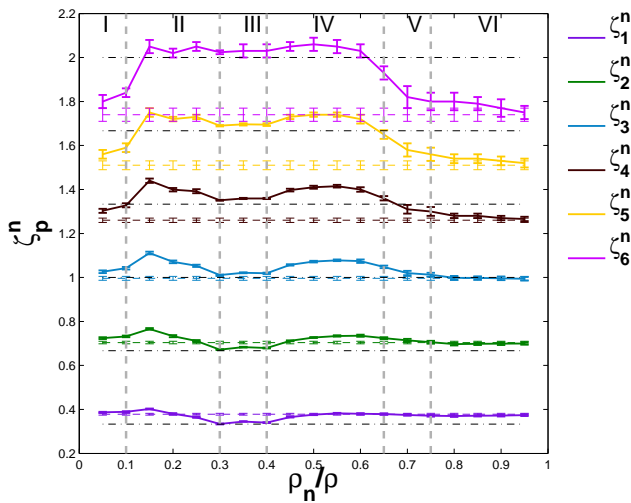}
}
\caption{\small Plots of $\zeta_p^n$, for $p=1$ to $6$, versus $\rho_n/\rho$,
from our shell-model runs $\tt B1$-$\tt B19$.  For the purpose of reference, we
show the value of a classical-fluid-turbulence exponent $\zeta^c_p$, for order $p$, by a
horizontal, dashed line; different colors indicate different values of the
order $p$.  The black, dot-dashed lines indicates $\zeta^{K41}_p=p/3$.  In the
shell-model runs $\tt B1$-$\tt B19$, we keep the mutual-friction coefficient
$B=1.5$ fixed.} 
\label{fig:zetadndBfixed} 
\end{figure}

We now examine the dependence of the multiscaling exponents $\zeta^i_p$, $i\in
(n,s)$ on $\rho_n/\rho$, while keeping the coefficient of mutual friction
$B=1.5$ fixed, in runs $\tt B1$-$\tt B19$. These runs allow us to classify the
behavior of $\zeta^i_p$, $i\in (n,s)$, as a function of $\rho_n/\rho$, more
clearly than the runs $\tt G1$-$\tt G9$. In the Supplemental Material, in
Table~\ref{table:zetapBfixed} we list the values of $\zeta^i_p$, $i\in (n,s)$,
which we extract from $\Sigma^i_p$ (Eq.~\ref{eq:strsigma}), for $p=1$ to $6$,
$i\in (n,s)$; each row of this Table has two lines; the first and second lines
contain the values of $\zeta^n_p$ and $\zeta^s_p$, respectively.  For these
runs $\zeta^n_p\simeq\zeta^s_p$. In Fig.~\ref{fig:zetadndBfixed} we plot $\zeta^n_p$,
versus $\rho_n/\rho$, for $p=1$ to $6$ in 
runs $\tt B1$-$\tt B19$.  These plots
show two regions ($0.1<\rho_n/\rho<0.3$ and $0.4<\rho_n/\rho<0.65$) with clear
bumps, where the values of $\zeta^n_p$ deviate significantly from both
$\zeta^{K41}_p(<\zeta^n_p)$ and $\zeta^c_p(<\zeta^n_p$). We classify
{\textit{roughly}} the behaviors of these $\zeta^n_p$ into six regions ${\rm
I}$-${\rm VI}$ (demarcated by grey, dashed, vertical lines in
Fig.~\ref{fig:zetadndBfixed}), which we describe below.  Region ${\rm I}$
($\rho_n/\rho\lesssim 0.1$): $\zeta^n_p\simeq\zeta^c_p$.  Region ${\rm II}$
($0.1<\rho_n/\rho< 0.3$): $\zeta^n_p$ differs significantly from both
$\zeta^c_p$ and $\zeta^{K41}_p$, with $\zeta^c_p<\zeta^n_p$ and
$\zeta^{K41}_p<\zeta^n_p$.  Region ${\rm III}$ ($0.3\lesssim\rho_n/\rho\lesssim
0.4$): $\zeta^n_p\simeq\zeta^{K41}_p$.  Region ${\rm IV}$ ($0.4<\rho_n/\rho<
0.65$): $\zeta^n_p$ differs significantly from both $\zeta^c_p$ and
$\zeta^{K41}_p$, with $\zeta^c_p<\zeta^n_p$ and $\zeta^{K41}_p<\zeta^n_p$.
Region ${\rm V}$ ($0.65\gtrsim\rho_n/\rho<0.75$): $\zeta^n_p$ shows a tendency
to move towards $\zeta^c_n$.  Region ${\rm VI}$ ($\rho_n/\rho\gtrsim0.75$):
$\zeta^n_p\simeq\zeta^c_p$.

\begin{figure}
\resizebox{\linewidth}{!}{
\includegraphics[height=12.cm]{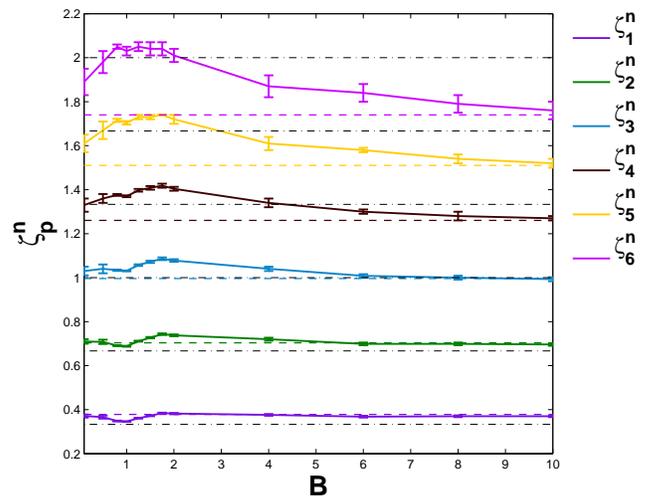}
}
\caption{\small Plot of $\zeta_p^n$, for $p=1$ to $6$, versus $\rho_n/\rho$,
from the shell-model runs $\tt R1$-$\tt R12$.  For the purpose of reference, we
show the value of a classical-fluid-turbulence exponent $\zeta^c_p$, for order $p$, by a
horizontal, dashed line; different colors indicate different values of the
order $p$.  The black, dot-dashed lines indicates $\zeta^{K41}_p=p/3$.  In the
shell-model runs $\tt R1$-$\tt R12$, we keep the normal-fluid density fraction
$\rho_n/\rho=0.5$ fixed.}
\label{fig:zetaBdndBfix} 
\end{figure}

We also explore the dependence of the multiscaling exponents $\zeta^i_p$, $i\in
(n,s)$, on the mutual-friction coefficient $B$,  while keeping the
normal-fluid-density fraction $\rho_n/\rho=0.5$ fixed.  In our
3D-HVBK-shell-model runs $\tt R1$-$\tt R12$, we systematically vary the values
of $B$; we list the values of $\zeta^i_p$, $i\in (n,s)$ obtained from
$\Sigma^i_p$ (Eq.~\ref{eq:strsigma}), for $p=1$ to $6$, in
Table~\ref{table:zetaprhonfixed} in the Supplemental Material; each row of this
Table has two lines; the first and second lines contain, respectively, the
values of $\zeta^n_p$ and $\zeta^s_p$.  In Fig.~\ref{fig:zetaBdndBfix} we plot
 $\zeta^n_p$ versus $B$, for $p=1$ to $6$, for the runs
$\tt R1$-$\tt R12$; the exponents $\zeta^n_p$ deviate significantly from their
classical-fluid-turbulence counterparts $\zeta^c_p$, in the range $1\leq B\leq3$, with
$\zeta^n_p>\zeta^c_p$, for $p\geq3$, $\zeta^n_1<\zeta^c_1$, and $\zeta^n_2$
marginally larger than $\zeta^c_2$.  As $B\rightarrow 0.1$ (small values) and
$B\rightarrow 10$ (large values) the multiscaling exponents
$\zeta^n_p\simeq\zeta^c_p$, because, in the limit $B\rightarrow 0$, the normal
fluid and superfluid are uncoupled; and for very large values of $B$, 
the coupling is so strong that single-fluid-turbulence results emerge. 

\begin{figure}
\resizebox{\linewidth}{!}{
\includegraphics[height=6.cm]{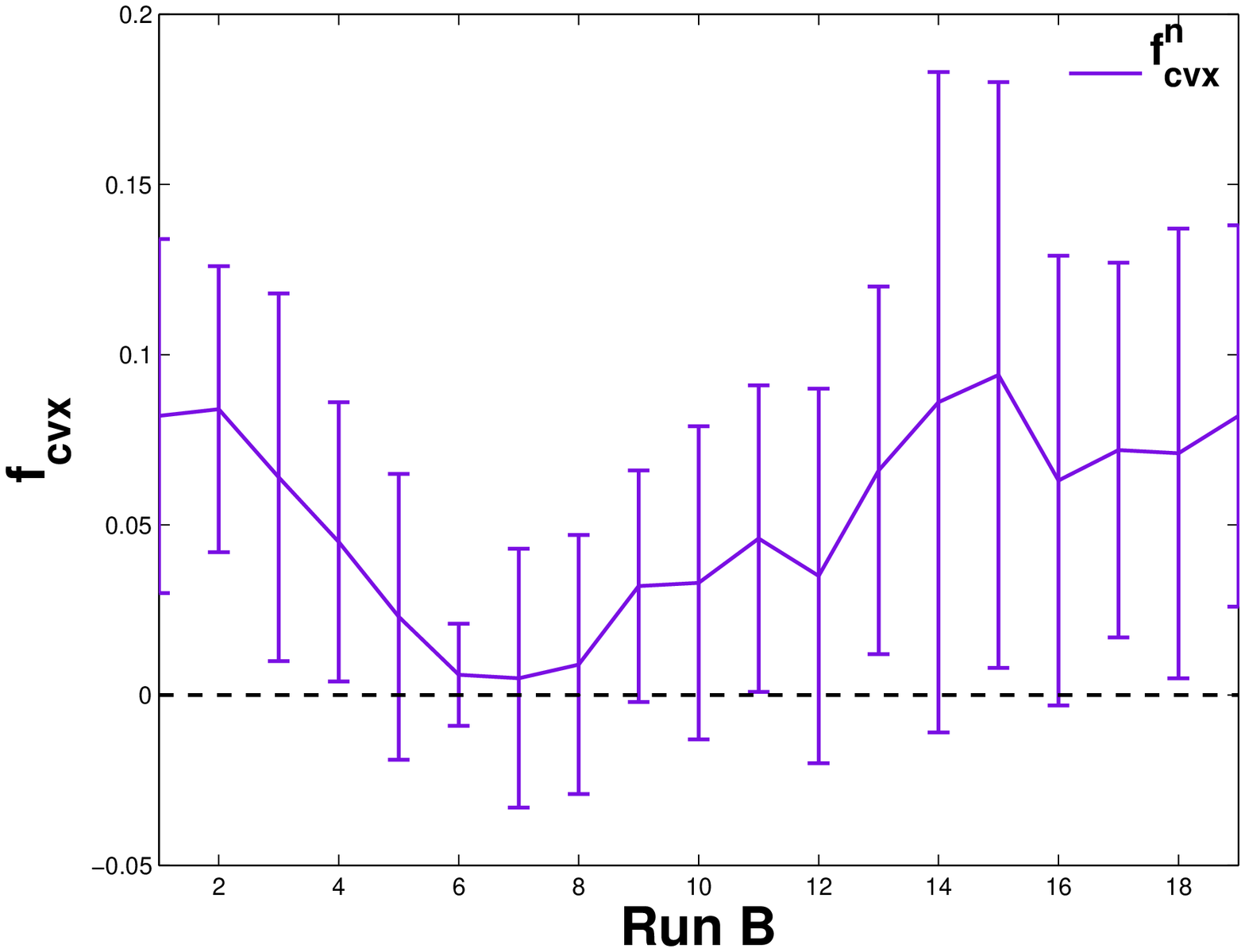}
\put(-100,30){\bf (a)}
\includegraphics[height=6.cm]{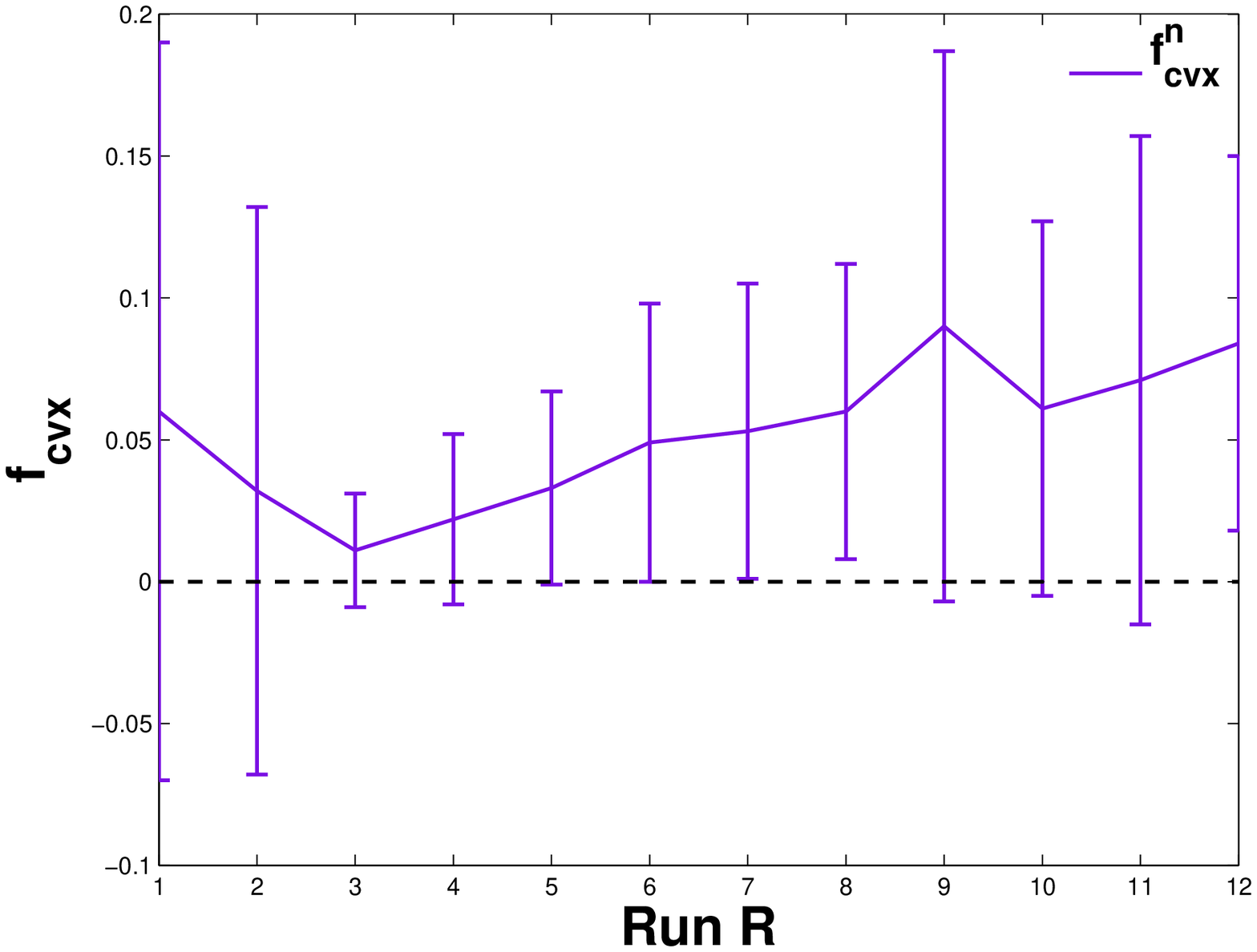}
\put(-100,30){\bf (b)}
}
\caption{\small Plots of (a) $f^n_{\rm cvx}$ for the runs $\tt B1$-$\tt B19$
($B=1.5$); (b) $f^n_{\rm cvx}$ for the runs $\tt R1$-$\tt R12$ ($\rho_n/\rho =
0.5$), where $f^i_{\rm
cvx}=(p_3-p_1)\zeta^i_{2p_2}-(p_3-p_2)\zeta^i_{2p_1}-(p_2-p_1)\zeta^i_{2p_3}$,
$i\in (n,s)$, and we take $p_1=1$, $p_2=2$, and $p_3=3$.  The multiscaling
exponents $\zeta^i_p$, $i\in (n,s)$, satisfy the convexity constraint, if
$f^i_{\rm cvx}>0$, for any three positive integers $p_1\leq p_2\leq p_3$.  The
$x$-axis label in the above plots indicates the run index, e.g., $\tt B1$.  } 
\label{fig:convexineq} 
\end{figure}

We have checked explicitly that all the values of $\zeta^n_p$ and $\zeta^s_p$,
which we have reported above, satisfy the convexity inequality
Eq.~(\ref{eq:convexchap6}).  We illustrate this in the plots of
Fig.~\ref{fig:convexineq}.

\section{Conclusions}
\label{sec:conclusions:ch6}

We have carried out extensive numerical simulations of the 3D-HVBK shell-model,
specifically to study the multiscaling of structure functions in superfluid
turbulence, because such multiscaling has been studied much less than its
counterpart in classical-fluid turbulence.  Experimental investigations of
turbulence in liquid helium, below the superfluid transition temperature
$T_{\lambda}$, have provided evidence for multiscaling, in the inertial
range~\cite{maurer1998local,salort2011investigation,salort2012fourfifthlaw}.
These experiments have also motivated our study. Direct numerical simulations
of models for superfluids, e.g., the Gross-Pitaevskii equation and the HVBK
two-fluid equations, have not been able to cover the large range of length
scales that are required to obtain reliable data for high-order structure
functions.  Shell models, based on the HVBK two-fluid equations, have been used
to study the statistical properties of 3D superfluid turbulence in both
$^4$He~\cite{wacks2011shellmodel,procacciaintermittencyshellmodel} and
$^3$He-B~\cite{wacks2011shellmodel,procacciashell3He}; these studies have
elucidated the natures of energy spectra and fluxes, for both forced,
statistically steady and decaying superfluid turbulence. The only detailed
investigation of the multiscaling behavior of structure functions is an
HVBK-shell-model study~\cite{procacciaintermittencyshellmodel}. This study has
shown that, for $\rho_n/\rho\leq0.1$ and $\rho_n/\rho\leq0.9$, the multiscaling
exponents are close to those in classical-fluid turbulence; whereas, in the
range $0.25\leq\rho_n/\rho\leq0.5$, high-order mutliscaling exponents deviate
significantly from, and are smaller than, their classical-fluid-turbulence
counterparts.

Our extensive study of the 3D-HVBK shell model has shown that the multiscaling
of structure functions in superfluid turbulence is more complex than that reported
in Ref.~\cite{procacciaintermittencyshellmodel}. However, our results agree
with those of Ref.~\cite{procacciaintermittencyshellmodel} in that, for
$\rho_n/\rho\lesssim 0.1$ and $\rho_n/\rho\gtrsim 0.75$, the multiscaling
exponents are close to the classical-fluid-turbulence values. Moreover, we find that there are
two regions, with $0.1<\rho_n/\rho< 0.3$ and $0.4<\rho_n/\rho< 0.65$, where the
multiscaling exponents are larger than their classical-fluid-turbulence and K41 counterparts,
i.e., $\zeta^i_p>\zeta^c_p$ and $\zeta^i_p>\zeta^{K41}_p$, $i\in(n,s)$. In the
range $0.3\lesssim\rho_n/\rho\lesssim 0.4$, these exponents are close to the
K41 prediction, i.e., $\zeta^i_p\simeq\zeta^{K41}_p$.  We have also
investigated the dependence of the multiscaling exponents on the
mutual-friction coefficient $B$, with $\rho_n/\rho=0.5$ fixed; our
results show that, for small (weak-coupling limit) and large (strong-coupling
limit) values of $B$, the multiscaling exponents tend to their classical-fluid-turbulence
values, whereas, in the range $1\lesssim B\lesssim3$, there are deviations from
the classical-fluid-turbulence behavior $\zeta^i_p>\zeta^c_p$, for $p\geq3$.  We
hope our extensive study of the multiscaling of structure functions in the
3D-HVBK shell-model will stimulate detailed experimental and DNS studies of
such multiscaling in quantum-fluid turbulence in different quantum fluids.

\vspace{0.25cm}
{\small\bf{ACKNOWLEDGMENTS}}
\vspace{0.25cm}

We thank CSIR, UGC, DST (India) and the Indo-French Centre for Applied
Mathematics (IFCAM) for financial support, and SERC (IISc) for computational
resources. 
VS acknowledges support from Centre Franco-Indien pour la Promotion de la
Recherche Avanc\'ee (CEFIPRA) project no. 4904.
We are grateful to A. Basu and S.S. Ray for useful discussions.

\bibliographystyle{apsrev4-1}
\bibliography{references}
\clearpage
\vspace{0.5cm}
{\large\bf{SUPPLEMENTAL MATERIAL}}
\vspace{0.5cm}

\begin{table*}
\begin{center}
\small
\resizebox{\linewidth}{!}{
   \begin{tabular}{@{\extracolsep{\fill}} c c c c c c c c c c c c c }
    \hline
    $ $ & $\rho_n/\rho$ &  $\zeta^n_1$  & $\zeta^n_2$ & $\zeta^n_3$ & $\zeta^n_4$ & $\zeta^n_5$ & $\zeta^n_6$\\ 
    $ $ & $(B)$ &  $\zeta^s_1$  & $\zeta^s_2$ & $\zeta^s_3$ & $\zeta^s_4$ & $\zeta^s_5$ & $\zeta^s_6$\\
   \hline \hline
    {\tt PG1}  &  $-$ &  $0.378\pm0.004$ & 
    $0.704\pm0.006$ & $0.996\pm0.009$ & $1.26\pm0.01$ & $1.51\pm0.02$ & $1.74\pm0.03$\\
    {\tt PG2}  &  $-$ &  $0.383\pm0.003$ & 
    $0.714\pm0.005$ & $1.007\pm0.007$ & $1.27\pm0.01$ & $1.52\pm0.02$ & $1.75\pm0.02$\\
   \hline \hline
    {\tt G1a}  &  $0.0450$ &  $0.378\pm 0.008$ & 
    $0.70\pm 0.01$ & $0.99\pm0.02$ & $1.26\pm0.03$ & $1.50\pm0.05$ & $1.73\pm0.07$\\
    {\tt }  &  $(1.5260)$ &  $0.378\pm 0.008$ & 
    $0.70\pm 0.01$ & $0.99\pm0.02$ & $1.26\pm0.03$ & $1.50\pm0.05$ & $1.73\pm0.07$\\
    \hline
    {\tt G1}  &  $0.0450$ &  $0.387\pm0.003$ & 
    $0.723\pm0.004$ & $1.024\pm0.006$ & $1.300\pm0.008$ & $1.56\pm0.01$ & $1.80\pm0.02$\\
    {\tt }  &  $(1.5260)$ &  $0.384\pm0.003$ & 
    $0.721\pm0.004$ & $1.022\pm0.006$ & $1.300\pm0.008$ & $1.55\pm0.01$ & $1.80\pm0.02$\\
    \hline
    {\tt G2}  &  $0.0998$ &  $0.390\pm0.003$ & 
    $0.732\pm0.005$ & $1.040\pm0.007$ & $1.32\pm0.01$ & $1.58\pm0.02$ & $1.83\pm0.03$\\
    {\tt }  &  $(1.3255)$ &  $0.389\pm0.003$ & 
    $0.731\pm0.005$ & $1.040\pm0.007$ & $1.32\pm0.01$ & $1.58\pm0.02$ & $1.83\pm0.03$\\
    \hline
    {\tt G3}  &  $0.2503$ &  $0.372\pm0.006$ & 
    $0.71\pm0.01$ & $1.02\pm0.01$ & $1.32\pm0.02$ & $1.60\pm0.02$ & $1.86\pm0.04$\\
    {\tt }  &  $(1.0765)$ &  $0.372\pm0.006$ & 
    $0.71\pm0.01$ & $1.02\pm0.01$ & $1.32\pm0.02$ & $1.60\pm0.02$ & $1.86\pm0.04$\\
    \hline
    {\tt G4}  &  $0.4004$ &  $0.3309\pm0.0001$ & 
    $0.6663\pm0.0001$ & $1.0046\pm0.0001$ & $1.3446\pm0.0001$ & $1.6858\pm0.0002$ & $2.0276\pm0.0003$\\
    {\tt }  &  $(0.9838)$ &  $0.3310\pm0.0001$ & 
    $0.6664\pm0.0001$ & $1.0044\pm0.0001$ & $1.3441\pm0.0001$ & $1.6847\pm0.0002$ & $2.0259\pm0.0002$\\
    \hline
    {\tt G5}  &  $0.4994$ &  $0.375\pm0.004$ & 
    $0.732\pm0.005$ & $1.079\pm0.006$ & $1.418\pm0.008$ & $1.75\pm0.01$ & $2.06\pm0.02$\\
    {\tt }  &  $(0.9848)$ &  $0.374\pm0.003$ & 
    $0.732\pm0.005$ & $1.079\pm0.006$ & $1.417\pm0.008$ & $1.74\pm0.01$ & $2.06\pm0.02$\\
    \hline
    {\tt G6}  &  $0.6003$ &  $0.386\pm0.003$ & 
    $0.737\pm0.005$ & $1.065\pm0.007$ & $1.37\pm0.01$ & $1.67\pm0.02$ & $1.94\pm0.03$\\
    {\tt }  &  $(1.0447)$ &  $0.385\pm0.003$ & 
    $0.737\pm0.005$ & $1.064\pm0.007$ & $1.37\pm0.01$ & $1.66\pm0.02$ & $1.94\pm0.03$\\
    \hline
    {\tt G7}  &  $0.6493$ &  $0.384\pm0.003$ & 
    $0.734\pm0.005$ & $1.060\pm0.006$ & $1.37\pm0.01$ & $1.66\pm0.02$ & $1.94\pm0.03$\\
    {\tt }  &  $(1.1034)$ &  $0.384\pm0.003$ & 
    $0.734\pm0.004$ & $1.060\pm0.006$ & $1.37\pm0.01$ & $1.66\pm0.02$ & $1.94\pm0.03$\\
    \hline
    {\tt G8}  &  $0.6995$ &  $0.383\pm0.003$ & 
    $0.728\pm0.004$ & $1.047\pm0.007$ & $1.35\pm0.01$ & $1.63\pm0.02$ & $1.90\pm0.03$\\
    {\tt }  &  $(1.1924)$ &  $0.383\pm0.003$ & 
    $0.728\pm0.004$ & $1.046\pm0.007$ & $1.35\pm0.01$ & $1.63\pm0.02$ & $1.90\pm0.03$\\
    \hline
    {\tt G9}  &  $0.7501$ &  $0.381\pm0.004$ & 
    $0.718\pm0.006$ & $1.027\pm0.008$ & $1.32\pm0.01$ & $1.59\pm0.02$ & $1.85\pm0.03$\\
    {\tt }  &  $(1.3267)$ &  $0.380\pm0.004$ & 
    $0.718\pm0.006$ & $1.027\pm0.008$ & $1.32\pm0.01$ & $1.59\pm0.02$ & $1.85\pm0.03$\\
    \hline
\hline
\end{tabular}
}
\end{center}
\caption{\small Multiscaling exponents $\zeta_p$ from our shell-model runs $\tt
PG1$, $\tt PG2$, and $\tt G1-G9$; each row of the Table has two lines; the
first and second lines contain, respectively, the values of $\zeta^n_p$ and
$\zeta^s_p$.  In the second column, $\rho_n/\rho$ is the normal-fluid density
fraction (first line) and $B$ is the mutual-friction coefficient (second line,
in parentheses).}
\label{table:zetapGruns}
\end{table*}

\begin{table*}
\begin{center}
\small
   \begin{tabular}{@{\extracolsep{\fill}} c c c c c c c c c c c c c }
    \hline
    $ $ & $\rho_n/\rho$ &  $\zeta_1$  & $\zeta_2$ & $\zeta_3$ & $\zeta_4$ & $\zeta_5$ & $\zeta_6$\\ 
   \hline \hline

    {\tt B1}  &  $0.05$ &  $0.387\pm0.002$ & 
    $0.724\pm0.004$ & $1.026\pm0.006$ & $1.303\pm0.009$ & $1.56\pm0.02$ & $1.80\pm0.03$\\
    {\tt }  &  $$ &  $0.384\pm0.002$ & 
    $0.720\pm0.004$ & $1.023\pm0.006$ & $1.301\pm0.009$ & $1.56\pm0.02$ & $1.80\pm0.03$\\

    {\tt B2}  &  $0.10$ &  $0.389\pm0.003$ & 
    $0.732\pm0.004$ & $1.042\pm0.006$ & $1.328\pm0.009$ & $1.59\pm0.02$ & $1.84\pm0.02$\\
    {\tt }  &  $$ &  $0.388\pm0.003$ & 
    $0.733\pm0.004$ & $1.045\pm0.005$ & $1.333\pm0.008$ & $1.60\pm0.01$ & $1.85\pm0.02$\\

    {\tt B3}  &  $0.15$ &  $0.402\pm0.002$ & 
    $0.766\pm0.004$ & $1.111\pm0.006$ & $1.44\pm0.01$ & $1.75\pm0.02$ & $2.05\pm0.03$\\
    {\tt }  &  $$ &  $0.395\pm0.002$ & 
    $0.761\pm0.004$ & $1.107\pm0.006$ & $1.44\pm0.01$ & $1.75\pm0.02$ & $2.05\pm0.03$\\

    {\tt B4}  &  $0.20$ &  $0.380\pm0.003$ & 
    $0.733\pm0.005$ & $1.071\pm0.006$ & $1.399\pm0.008$ & $1.72\pm0.01$ & $2.02\pm0.02$\\
    {\tt }  &  $$ &  $0.376\pm0.003$ & 
    $0.730\pm0.005$ & $1.069\pm0.006$ & $1.397\pm0.008$ & $1.71\pm0.01$ & $2.02\pm0.02$\\

    {\tt B5}  &  $0.025$ &  $0.364\pm0.002$ & 
    $0.711\pm0.004$ & $1.053\pm0.005$ & $1.392\pm0.009$ & $1.73\pm0.01$ & $2.05\pm0.02$\\
    {\tt }  &  $$ &  $0.360\pm0.002$ & 
    $0.708\pm0.004$ & $1.051\pm0.005$ & $1.390\pm0.007$ & $1.72\pm0.01$ & $2.05\pm0.02$\\

    {\tt B6}  &  $0.30$ &  $0.334\pm0.002$ & 
    $0.672\pm0.002$ & $1.011\pm0.002$ & $1.351\pm0.002$ & $1.690\pm0.003$ & $2.024\pm0.09$\\
    {\tt }  &  $$ &  $0.334\pm0.002$ & 
    $0.671\pm0.002$ & $1.010\pm0.002$ & $1.350\pm0.002$ & $1.688\pm0.003$ & $2.01\pm0.009$\\

    {\tt B7}  &  $0.35$ &  $0.345\pm0.001$ & 
    $0.683\pm0.002$ & $1.021\pm0.002$ & $1.359\pm0.003$ & $1.696\pm0.005$ & $2.03\pm0.01$\\
    {\tt }  &  $$ &  $0.3385\pm0.0009$ & 
    $0.677\pm0.001$ & $1.015\pm0.002$ & $1.355\pm0.003$ & $1.693\pm0.005$ & $2.02\pm0.01$\\

    {\tt B8}  &  $0.40$ &  $0.340\pm0.002$ & 
    $0.679\pm0.002$ & $1.019\pm0.003$ & $1.359\pm0.003$ & $1.695\pm0.006$ & $2.03\pm0.01$\\
    {\tt }  &  $$ &  $0.339\pm0.001$ & 
    $0.680\pm0.002$ & $1.021\pm0.002$ & $1.361\pm0.002$ & $1.699\pm0.006$ & $2.03\pm0.01$\\

    {\tt B9}  &  $0.45$ &  $0.365\pm0.002$ & 
    $0.712\pm0.002$ & $1.057\pm0.003$ & $1.397\pm0.006$ & $1.73\pm0.01$ & $2.05\pm0.02$\\
    {\tt }  &  $$ &  $0.353\pm0.001$ & 
    $0.699\pm0.002$ & $1.046\pm0.003$ & $1.389\pm0.005$ & $1.72\pm0.01$ & $2.05\pm0.03$\\

    {\tt B10}  &  $0.50$ &  $0.376\pm0.002$ & 
    $0.727\pm0.002$ & $1.072\pm0.004$ & $1.410\pm0.007$ & $1.74\pm0.01$ & $2.06\pm0.03$\\
    {\tt }  &  $$ &  $0.365\pm0.002$ & 
    $0.720\pm0.002$ & $1.068\pm0.004$ & $1.408\pm0.007$ & $1.74\pm0.01$ & $2.05\pm0.03$\\

    {\tt B11}  &  $0.55$ &  $0.382\pm0.002$ & 
    $0.734\pm0.003$ & $1.078\pm0.004$ & $1.415\pm0.006$ & $1.74\pm0.01$ & $2.05\pm0.03$\\
    {\tt }  &  $$ &  $0.370\pm0.002$ & 
    $0.727\pm0.003$ & $1.075\pm0.004$ & $1.414\pm0.007$ & $1.74\pm0.01$ & $2.06\pm0.03$\\

    {\tt B12}  &  $0.60$ &  $0.380\pm0.003$ & 
    $0.735\pm0.005$ & $1.074\pm0.008$ & $1.40\pm0.01$ & $1.72\pm0.02$ & $2.03\pm0.03$\\
    {\tt }  &  $$ &  $0.379\pm0.003$ & 
    $0.734\pm0.005$ & $1.073\pm0.008$ & $1.40\pm0.01$ & $1.72\pm0.02$ & $2.03\pm0.03$\\

    {\tt B13}  &  $0.65$ &  $0.379\pm0.003$ & 
    $0.724\pm0.004$ & $1.048\pm0.006$ & $1.36\pm0.01$ & $1.65\pm0.02$ & $1.93\pm0.03$\\
    {\tt }  &  $$ &  $0.378\pm0.003$ & 
    $0.723\pm0.004$ & $1.047\pm0.006$ & $1.36\pm0.01$ & $1.65\pm0.02$ & $1.93\pm0.03$\\

    {\tt B14}  &  $0.70$ &  $0.375\pm0.004$ & 
    $0.714\pm0.007$ & $1.02\pm0.01$ & $1.31\pm0.02$ & $1.58\pm0.03$ & $1.82\pm0.05$\\
    {\tt }  &  $$ &  $0.375\pm0.005$ & 
    $0.714\pm0.007$ & $1.02\pm0.01$ & $1.31\pm0.02$ & $1.58\pm0.03$ & $1.82\pm0.05$\\

    {\tt B15}  &  $0.75$ &  $0.371\pm0.004$ & 
    $0.706\pm0.006$ & $1.012\pm0.009$ & $1.30\pm0.02$ & $1.56\pm0.03$ & $1.80\pm0.04$\\
    {\tt }  &  $$ &  $0.371\pm0.004$ & 
    $0.705\pm0.006$ & $1.012\pm0.009$ & $1.30\pm0.02$ & $1.56\pm0.03$ & $1.80\pm0.04$\\

    {\tt B16}  &  $0.80$ &  $0.370\pm0.004$ & 
    $0.697\pm0.006$ & $0.998\pm0.009$ & $1.28\pm0.01$ & $1.54\pm0.02$ & $1.80\pm0.04$\\
    {\tt }  &  $$ &  $0.369\pm0.004$ & 
    $0.697\pm0.006$ & $0.998\pm0.009$ & $1.28\pm0.01$ & $1.54\pm0.02$ & $1.80\pm0.04$\\

    {\tt B17}  &  $0.85$ &  $0.371\pm0.003$ & 
    $0.698\pm0.005$ & $0.998\pm0.007$ & $1.28\pm0.01$ & $1.54\pm0.02$ & $1.79\pm0.03$\\
    {\tt }  &  $$ &  $0.371\pm0.003$ & 
    $0.698\pm0.005$ & $0.998\pm0.007$ & $1.28\pm0.01$ & $1.54\pm0.02$ & $1.79\pm0.03$\\

    {\tt B18}  &  $0.90$ &  $0.372\pm0.004$ & 
    $0.699\pm0.006$ & $0.996\pm0.008$ & $1.27\pm0.01$ & $1.53\pm0.02$ & $1.77\pm0.04$\\
    {\tt }  &  $$ &  $0.372\pm0.004$ & 
    $0.698\pm0.006$ & $0.996\pm0.008$ & $1.27\pm0.01$ & $1.53\pm0.02$ & $1.77\pm0.04$\\

    {\tt B19}  &  $0.95$ &  $0.374\pm0.004$ & 
    $0.700\pm0.006$ & $0.994\pm0.008$ & $1.266\pm0.01$ & $1.52\pm0.02$ & $1.75\pm0.03$\\
    {\tt }  &  $ $ &  $0.374\pm0.004$ & 
    $0.699\pm0.006$ & $0.994\pm0.008$ & $1.265\pm0.01$ & $1.52\pm0.02$ & $1.75\pm0.03$\\
\hline
\end{tabular}
\end{center}
\caption{\small Multiscaling exponents $\zeta_p$ from our shell-model runs $\tt
B1$-$\tt B19$; each row of the Table has two lines; the first and second lines
contain, respectively, the values of $\zeta^n_p$ and $\zeta^s_p$. $\rho_n/\rho$
is the normal-fluid density fraction; we keep the mutual-friction coefficient
$B=1.5$ fixed.  }
\label{table:zetapBfixed}
\end{table*}

\begin{table*}
\begin{center}
\small
   \begin{tabular}{@{\extracolsep{\fill}} c c c c c c c c c c c c c }
    \hline
    $ $ & $B$ &  $\zeta_1$  & $\zeta_2$ & $\zeta_3$ & $\zeta_4$ & $\zeta_5$ & $\zeta_6$\\ 
   \hline \hline

    {\tt R1}  &  $0.10$ &  $0.371\pm0.007$ & 
    $0.71\pm0.01$ & $1.03\pm0.02$ & $1.33\pm0.03$ & $1.61\pm0.04$ & $1.89\pm0.06$\\
    {\tt }  &  $$ &  $0.370\pm0.007$ & 
    $0.71\pm0.01$ & $1.03\pm0.02$ & $1.33\pm0.03$ & $1.61\pm0.04$ & $1.89\pm0.06$\\

    {\tt R2}  &  $0.50$ &  $0.366\pm0.007$ & 
    $0.708\pm0.01$ & $1.04\pm0.02$ & $1.36\pm0.02$ & $1.67\pm0.04$ & $1.98\pm0.05$\\
    {\tt }  &  $$ &  $0.366\pm0.007$ & 
    $0.708\pm0.01$ & $1.04\pm0.02$ & $1.36\pm0.02$ & $1.67\pm0.04$ & $1.98\pm0.05$\\

    {\tt R3}  &  $0.80$ &  $0.348\pm0.002$ & 
    $0.691\pm0.002$ & $1.034\pm0.003$ & $1.376\pm0.004$ & $1.715\pm0.007$ & $2.05\pm0.01$\\
    {\tt }  &  $$ &  $0.347\pm0.002$ & 
    $0.691\pm0.002$ & $1.033\pm0.003$ & $1.374\pm0.004$ & $1.712\pm0.007$ & $2.04\pm0.01$\\

    {\tt R4}  &  $1.00$ &  $0.346\pm0.001$ & 
    $0.688\pm0.002$ & $1.031\pm0.002$ & $1.370\pm0.004$ & $1.704\pm0.008$ & $2.03\pm0.02$\\
    {\tt }  &  $$ &  $0.345\pm0.001$ & 
    $0.688\pm0.002$ & $1.030\pm0.002$ & $1.368\pm0.004$ & $1.700\pm0.008$ & $2.02\pm0.02$\\

    {\tt R5}  &  $1.25$ &  $0.361\pm0.002$ & 
    $0.711\pm0.002$ & $1.057\pm0.004$ & $1.397\pm0.006$ & $1.73\pm0.01$ & $2.05\pm0.02$\\
    {\tt }  &  $$ &  $0.359\pm0.002$ & 
    $0.710\pm0.002$ & $1.056\pm0.004$ & $1.394\pm0.006$ & $1.72\pm0.01$ & $2.04\pm0.02$\\

    {\tt R6}  &  $1.50$ &  $0.372\pm0.002$ & 
    $0.727\pm0.003$ & $1.072\pm0.005$ & $1.408\pm0.008$ & $1.73\pm0.01$ & $2.04\pm0.03$\\
    {\tt }  &  $$ &  $0.370\pm0.002$ & 
    $0.725\pm0.003$ & $1.070\pm0.005$ & $1.405\pm0.008$ & $1.73\pm0.01$ & $2.04\pm0.03$\\

    {\tt R7}  &  $1.75$ &  $0.384\pm0.002$ & 
    $0.743\pm0.004$ & $1.086\pm0.006$ & $1.418\pm0.009$ & $1.74\pm0.002$ & $2.04\pm0.03$\\
    {\tt }  &  $$ &  $0.382\pm0.002$ & 
    $0.742\pm0.004$ & $1.086\pm0.006$ & $1.417\pm0.009$ & $1.73\pm0.002$ & $2.04\pm0.03$\\

    {\tt R8}  &  $2.00$ &  $0.382\pm0.003$ & 
    $0.738\pm0.004$ & $1.078\pm0.006$ & $1.404\pm0.009$ & $1.72\pm0.02$ & $2.01\pm0.03$\\
    {\tt }  &  $$ &  $0.382\pm0.003$ & 
    $0.738\pm0.004$ & $1.077\pm0.006$ & $1.403\pm0.009$ & $1.73\pm0.02$ & $2.01\pm0.03$\\

    {\tt R9}  &  $4.00$ &  $0.376\pm0.004$ & 
    $0.720\pm0.007$ & $1.04\pm0.01$ & $1.34\pm0.02$ & $1.61\pm0.03$ & $1.87\pm0.05$\\
    {\tt }  &  $$ &  $0.376\pm0.004$ & 
    $0.720\pm0.007$ & $1.04\pm0.01$ & $1.34\pm0.02$ & $1.61\pm0.03$ & $1.87\pm0.05$\\

    {\tt R10}  &  $6.00$ &  $0.368\pm0.004$ & 
    $0.699\pm0.006$ & $1.008\pm0.008$ & $1.30\pm0.01$ & $1.58\pm0.02$ & $1.84\pm0.04$\\
    {\tt }  &  $$ &  $0.368\pm0.004$ & 
    $0.699\pm0.006$ & $1.008\pm0.008$ & $1.30\pm0.01$ & $1.58\pm0.02$ & $1.84\pm0.04$\\

    {\tt R11}  &  $8.00$ &  $0.370\pm0.004$ & 
    $0.699\pm0.006$ & $1.000\pm0.009$ & $1.28\pm0.02$ & $1.54\pm0.02$ & $1.79\pm0.04$\\
    {\tt }  &  $$ &  $0.370\pm0.004$ & 
    $0.699\pm0.006$ & $1.000\pm0.009$ & $1.28\pm0.01$ & $1.54\pm0.02$ & $1.79\pm0.04$\\

    {\tt R12}  &  $10.0$ &  $0.370\pm0.004$ & 
    $0.696\pm0.006$ & $0.993\pm0.009$ & $1.27\pm0.01$ & $1.52\pm0.02$ & $1.76\pm0.04$\\
    {\tt }  &  $$ &  $0.370\pm0.004$ & 
    $0.696\pm0.006$ & $0.992\pm0.009$ & $1.27\pm0.01$ & $1.52\pm0.02$ & $1.76\pm$0.04\\
\hline
\end{tabular}
\end{center}
\caption{\small Multiscaling exponents $\zeta_p$ from our shell-model runs $\tt
R1$-$\tt R12$; each row of the Table has two lines; the first and second lines
contain, respectively, the values of $\zeta^n_p$ and $\zeta^s_p$. $B$ is the
mutual-friction coefficient; we keep the normal-fluid-density fraction
$\rho_n/\rho=0.5$ fixed.  }
\label{table:zetaprhonfixed}
\end{table*}

\end{document}